\documentclass[preprint,aps,prb]{revtex4-1}
\usepackage[pdftex]{graphics}
\usepackage{epstopdf}
\usepackage{amsmath}
\usepackage{amssymb}
\usepackage{upgreek}
\usepackage{subfloat}
\usepackage{float}
\usepackage{dcolumn}
\usepackage{bm}
\usepackage{setspace}
\usepackage{color}
\usepackage{relsize}
\usepackage{graphicx}

\newcommand{\comment}[1]{}

\makeatletter

\begin{document}

\title{Optical Self Energy in Graphene due to Correlations}

\author{J. Hwang$^{1}$}
\email{jungseek@skku.edu; corresponding author}
\author{J. P. F. LeBlanc$^{2,3}$}
\author{J. P. Carbotte$^{4,5}$}
\affiliation{$^1$Department of Physics, Sungkyunkwan University, Suwon, Gyeonggi-do
440-746, Republic of Korea}
\affiliation{$^2$Department of Physics, University of Guelph,
Guelph, Ontario N1G 2W1 Canada}
\affiliation{$^3$Guelph-Waterloo Physics Institute, University of Guelph, Guelph, Ontario N1G 2W1 Canada}
\affiliation{$^4$Department of Physics and Astronomy, McMaster
University, Hamilton, Ontario L8S 4M1 Canada}
\affiliation{$^5$The Canadian Institute for Advanced Research, Toronto, ON M5G 1Z8 Canada}

\date{\today}

\begin{abstract}

In highly correlated systems one can define an optical self energy in analogy to its quasiparticle (QP) self energy counterpart. This quantity provides useful information on the nature of the excitations involved in  inelastic scattering processes. Here we calculate the self energy of the intraband optical transitions in graphene originating in the electron-electron interaction (EEI) as well as electron-phonon interaction (EPI). Although optics involves an average over all momenta ($k$) of the charge carriers, the structure in the optical self energy is nevertheless found to mirror mainly that of the corresponding quasiparticles for $k$ equal to or near the Fermi momentum $k_F$. Consequently plasmaronic structures which are associated with momenta near the Dirac point at $k=0$ are not important in the intraband optical response. While the structure of the electron-phonon interaction (EPI) reflects the sharp peaks of the phonon density of states, the excitation spectrum associated with the electron-electron interaction is in comparison structureless and flat and extends over an energy range which scales linearly with the value of the chemical potential. Modulations seen on the edge of the interband optical conductivity as it rises towards its universal background value are traced to structure in the quasiparticle self energies around $k_F$ of the lower Dirac cone associated with the occupied states.

\end{abstract}
\pacs{78.67.Wj,72.80.Vp}

\maketitle

\section{Introduction}

The isolation of a single layer of graphene\cite{novoselov:2005,novoselov:nature:2005,zhang:2005} lead directly to the establishment of the charge carriers in this system as effectively massless Dirac fermions. The carrier mobility in graphene is particularly large and the number of carriers can be modified through the gate voltage in a field effect configuration. It is also considered promising for the development of a new generation of electronic devices\cite{geim:2009}.  Several reviews\cite{geim:2007,neto:2009,gusynin:2007:ijmp,abergel:2010,orlita:2010} have appeared which document the unusual charge transport in this material.

The optical properties of graphene are particularly interesting and an experimental review was given by Orlita and Potemski\cite{orlita:2010}. To summarize, graphene shows a Drude-like intraband conductivity centered around $\Omega = 0$ which is followed at higher energies by a region of reduced conductivity before the sharp rise to a constant universal background value of $\sigma_0=\pi e^2/(2h)$\cite{li:2008,wang:2008,mak:2008,nair:2008,peres:2008,gusynin:2007,gusynin:2009,carbotte:2010} ($e$ is the charge on the electron and $h$ Planck's constant) at $\Omega$ equal to twice the value of the chemical potential $\mu$. This part of graphene's optical response is due to the interband transitions between the fully occupied lower Dirac cone of the valence band and the unoccupied part of the conduction band Dirac cone. Another interesting optical property is a giant Faraday rotation. As demonstrated by Crassee et al.\cite{crassee:2011a,crassee:2011}, a single layer of graphene turns the polarization of incident light by several degrees in relatively modest magnetic fields. This property could be useful for ultra thin infrared magneto-optical devices. A related effect is the enhanced dichroism of graphene nanoribbons\cite{hipolito:2011} where the polarization state of the light effects its absorption. While the effects described above, and many others, can be understood within a bare band picture, some cannot. As an example. a large amount of absorption, of the order of one third the universal background value $\sigma_0=\pi e^2/(2h)$ is observed in the Pauli blocked region of the spectrum which lies between the intraband Drude peak and the universal background which occurs above $\Omega \geq 2\mu$ due to interband transitions. This additional absorption can be understood to arise partially from the electron-phonon interactions (EPI)\cite{geim:2007,carbotte:2010} with an additional contribution from electron-electron (EEI) correlations\cite{grushin:2009}. These many-body effects provide a finite self energy part to the electronic spectral density, $A(k,\omega)$, for momentum, $k$, and energy ,$\omega$. This quantity is measured directly in angular resolved photoemission experiments (ARPES)\cite{bostwick:2007,bianchi:2010,bostwick:2010,zhou:2008,schachinger:2003} for the occupied values of $k$. Recent ARPES data on graphene has revealed the splitting of the Dirac point into two separate points with a region in between showing plasmaronic features associated with a coupling of electronic states with plasmons.  Such structures are also predicted to exist in the density of states, $N(\omega)$, which depends on an average of $A(k, \omega)$ over all momentum values, $\boldsymbol{k}$, but have yet to be clearly isolated from other many body effects in scanning tunnelling microscopy (STM)\cite{li:2009,miller:2009,brar:2010,leblanc:2011}. In this regard graphene is different from a conventional metal in which the renormalized density of states is not expected to show sharp many body signatures, in as much as its bare band value does not vary significantly on the energy scale of the exchanged boson involved.\cite{nicol:2009,mitrovic:1983,mitrovic:1983:2}However, the linear in $\omega$ dependence of $N(\omega)$ in graphene is sufficient to modify this expectation.

Of course there are many known effects of electron-electron (EEI) and electron-phonon (EPI) interactions which can provide profound modifications of the physical properties of metals. For example they lead to inelastic scattering\cite{schachinger:2000,carbotte:1995,nicol:1991,nicol:1991:2,schachinger:1997} and in the superconducting state they provide the so called strong coupling corrections\cite{carbotte:1986,marsiglio:1992,mitrovic:1980} to canonical Bardeen-Cooper-Schrieffer (BCS) theory. These corrections are remarkably well described by Eliashberg theory for electron-phonon coupling.
Inelastic scattering is quite distinct from other possible complications such as the presence of van-Hove singularities\cite{schachinger:1990,arberg:1993} and even strong scattering anisotropies\cite{branch:1995,leung:1976,odonovan:1995} since these complexities are not expected to be important in discussion of the low energy properties of graphene.

In this paper we calculated the effect of electron-electron interactions on the optical self energy \cite{carbotte:2011} of the massless Dirac fermions of graphene. The optical self energy is defined in direct analogy to the quasiparticle self energy. It is however a two body property and is defined in terms of a complex generalized Drude form in which both carrier effective mass and scattering rate acquire a temperature and frequency dependence. Optical mass renormalization and scattering rate are the central elements to be calculated here for the intraband transitions. For comparison with the EEI case, and to provide additional insight into that case, we also study the effect of the EPI on the same quantities. This involves the exchange of phonons and provides well known many-body corrections to Drude theory in conventional metals, such as the Holstein phonon assisted absorption sidebands\cite{marsiglio:1998,marsiglio:1999,jhwang:2008,mitrovic:1985,sharapov:2005}. In this case details of the electron-phonon spectral density denoted by $\alpha^2F(\omega)$ are encoded in the optical mass and scattering rate and these have been``inverted"\cite{carbotte:2011} to yield $\alpha^2F(\omega)$ which is a dimensionless function closely related to the phonon distribution $F(\omega)$. The difference between these two fundamental functions lies in that each phonon in $\alpha^2F(\omega)$ is further weighted by an appropriate factor related to the electron-phonon interaction strength\cite{marsiglio:1998,marsiglio:1999,jhwang:2008}. By contrast, for EEI it is not from the outset guaranteed that such a boson exchange concept can be applied to the excitations involved, be they particle-hole, plasmons or plasmarons. Nevertheless it will turn out to be helpful in the interpretation of the structure seen in optical properties when the electron-electron interactions are included, to understand how such a concept can be approximately applied to the results.

Along with the intraband optical transitions which provide a Drude-like respond center around $\omega = 0$ and which are analyzed in detail here, graphene also exhibits interband transitions. These set in only at photon energies, $\omega$, comparable to twice the value of the chemical potential, $2\mu$, and provide a nearly constant universal background $\sigma_0=\pi e^2/(2h)$. These optical transitions are not part of the usual single-band metallic case and cannot be naturally analyzed with an extended Drude form. Nevertheless we find that the structure seen in the low energy rising edge of the interband piece can also be related to well defined excitations of the electrons.

In section II we define the optical self energy, $\Sigma^{op}(\omega)$, and relate it to the optical conductivity which can be calculated from its Kubo formula as an appropriate overlap of two electronic spectral functions  for the same momentum but with energy arguments displaced by the photon energy, $\omega$. We give the formulas needed to calculate the quasiparticle self energy, $\Sigma^{qp}_s(k, \omega)$, which enters the spectral function. In general $\Sigma^{qp}_{s}(k, \omega)$ depends on momentum, $k$, energy, $\omega$, and on valley index, $s=+/-$ for conduction and valence band respectively. To include electron-electron interactions we employ a random phase approximation for the dynamical screening of the potential between electrons. This latter quantity is reduced by the medium dielectric constant which enters in its denominator as an average value of the materials above and below the graphene sheet. We present our numerical results for pure EEI, EPI and combined EEI+EPI, and begin by emphasizing the structure that enters the optical mass renormalization, $\lambda^{op}(\omega)$, as a function of $\omega$. This structure reflects details of the excitation spectrum to which the massless Dirac electrons are coupled. We compare the structure seen in $\lambda^{op}(\omega)$ with the structure that enters its counterpart, $\lambda^{qp}(k, \omega)$, which can depend on momentum as well as on $\omega$ and emphasize that only its value at or near $k=k_F$ appears to enter importantly into $\lambda^{op}(\omega)$. In section III we introduce simplified,  approximate but analytic formulas which apply to the case of a general energy dependent electronic density of state and specialize them to the specific case of graphene. We obtain formulas for real and imaginary part of the quasiparticle and optical self energy in a general boson exchange model and compare them. Analytic results are given for their zero frequency limits ($\omega=0$). Section IV deals with the problems of fitting quasiparticle and optical self energy data obtained from complete calculations based on the Kubo formula, to our simplified equations for a boson exchange mechanism. A maximum entropy inversion technique is used to obtain the best fitting electron-boson spectral density. The consistency of such an approach is checked, and its limitations commented on. Despite these limitation the procedure yields insight into the nature of the excitation spectrum that provides the scattering when electron-electron correlations are included. In section V we consider the case of the boson structure that arises in the interband part of the optical conductivity of graphene. Section VI contains our conclusions.

\section{Intraband Optical Self Energy}

It has become common to analyze the optical conductivity, $\sigma(T,\omega)$, of a correlated electron system in terms of a generalized Drude form in which an optical self energy, $\Sigma^{op}(T,\omega)$, is introduced.  Here, $T$ is temperature and $\omega$ is photon energy.  We write\cite{jhwang:2008}
\begin{equation}\label{eq1}
\sigma(T,\omega)\equiv\sigma_1(T,\omega)+i\sigma_2(T,\omega)=\frac{i}{4\pi}\frac{\Omega_p^2}{\omega-2\Sigma^{op}(T,\omega)}
\end{equation}
where $\sigma_1(T,\omega)$ and $\sigma_2(T,\omega)$ are the real and imaginary parts of the conductivity and $\Omega_p$ is the plasma energy. It is further instructive to write $\Sigma^{op}(T,\omega)=\Sigma_1^{op}+i\Sigma_2^{op}$ in terms of its real and imaginary part with $-2\Sigma_2^{op}(T,\omega) \equiv 1/\tau^{op}(T,\omega)$ and $-2 \Sigma_1^{op}(T,\omega)\equiv \omega [ m^*_{op}(T,\omega)/m -1]$ where $\tau^{op}(T,\omega)$ is an optical scattering time and $m^*_{op}(T,\omega)/m -1 \equiv \lambda^{op}(T,\omega)$ is an optical mass renormalization factor. In terms of these quantities the conductivity takes on its usual non-interacting form but now with frequency dependent scattering rate and effective mass. In particular, the real part of $\sigma(T, \omega)$ is
\begin{equation}\label{eq2}
\sigma_1(T,\omega)=\frac{\Omega_p^2}{4\pi}\frac{1/\tau^{op}(T,\omega)}{[\omega(1+\lambda^{op}(T,\omega))]^2
+(1/\tau^{op}(T, \omega))^2}.
\end{equation}
For graphene the Kubo formula gives the conductivity of the massless Dirac fermions in terms of their spectral density, $A^\pm(k,\omega)$,\cite{leblanc:2011} where $k$ is momentum and $\omega$ is energy.  The `$\pm$' notation refers to the conduction and valence bands respectively.  For the bare bands, $\epsilon_{\pm}(k)\equiv \pm v_F |k|$ with $v_F$ the Fermi velocity. These bands define two Dirac cones which meet at $k=0$, the Dirac point. In graphene there are two such points in the Brillouin zone at $K$ and $K^\prime$ points which provide a valley degeneracy factor, $g_v=2$, in addition to the standard spin degeneracy $g_s=2$.\cite{carbotte:2010}

The the real part of the conductivity is given by
\begin{equation}\label{eq3}
\frac{\sigma_1(T,\omega)}{\sigma_0}=\frac{4}{\omega}\int_{-\infty}^\infty d\omega^\prime \left[ f(\omega^\prime)-f(\omega^\prime+\omega) \right]
\int_{0}^{W_c} k dk A(k,\omega^\prime) A(k,\omega^\prime +\omega),
\end{equation}
where $W_C$ is the energy of the band cutoff. The imaginary part of the conductivity can be obtained by using a Kramers-Kronig relation.
\begin{equation}\label{eq3a}
\sigma_2(T, \omega)=-\frac{2\omega}{\pi} P \int^{\infty}_{0}\frac{\sigma_1(T, \omega)}{\omega'^2-\omega^2}d\omega'
\end{equation}
where $P$ stands for the principal value integration.

Here, the total charge carrier spectral function, $A(k,\omega)$ is given by\cite{leblanc:2011}
\begin{equation}\label{eq4}
A(k,\omega)=\sum_{s=\pm} \frac{1}{\pi}\frac{-{\rm Im}\Sigma_s^{qp}(k,\omega)}{[\omega-{\rm Re}\Sigma_s^{qp}(k,\omega)-\epsilon_k^s]^2+[{\rm Im}\Sigma_s^{qp}(k,\omega)]^2}
\end{equation}
where $\epsilon_k^s= sv_F |k| -\mu_0$ with $v_F$ the Fermi velocity, $\mu_0$ the chemical potential and $\Sigma^{qp}_s(k,\omega)$  the quasiparticle self energy due to many body interactions which can depend on the band index, $s$.
Here, for simplicity, we will be interested in treating the case of an electron-phonon interaction with coupling to a single Einstein mode at energy $\omega=\omega_E$.\cite{park:2007,park:2008,park:2009} Detailed calculations of the electronic self energies in graphene due to electron-phonon coupling have been done in density functional theory.\cite{park:2007,park:2008,park:2009} An important observation made in reference\cite{park:2007} for the present work is that the results of such complex computations have little dependence on the direction and magnitude of the electron momentum $\vec{k}$ and can be modeled in a first approximation through coupling to a single phonon mode at energy $\omega = 200$ meV. Here we adopt this model but will also allow for coupling (still assumed independent of direction and magnitude of $\vec{k}$) to a group of phonons rather than a single mode. This can be accomplished by the introduction of an electron-boson spectral density $\alpha^2F(\omega)$ again assumed to be independent of the Dirac fermion momentum $\vec{k}$. With such a simplified model, the electron self energy is momentum and band index independent and follows from the equation\cite{dogan:2003}
\begin{eqnarray}\label{eq5}
\Sigma^{EPI}(k,\omega)=\int_{0}^{\infty} \alpha^2 F(\nu)d\nu \int_{-\infty}^{\infty} d\omega^\prime \frac{N(\omega^\prime)}{N_0}\left[ \frac{\Theta(-\omega^\prime)}{\omega-\nu-\omega^\prime+i0^+}  +\frac{\Theta(\omega^\prime)}{\omega+\nu-\omega^\prime+i0^+}\right]
\end{eqnarray}
with $N(\omega^\prime)/N_0$ the density of electronic states which for graphene we take as $|\omega^\prime+\mu_0|/\mu_0$ as a first approximation and the Heaviside function $\Theta(\omega)$. There is no dependence of this self energy on the electron momentum, $k$.

The electron-electron interaction provides a second contribution to the self energy \cite{wunsch:2006, polini:2008, hwang:2008, sensarma:2011, barlas:2007, hwang:2007} of the massless Dirac quasiparticles in graphene which is more complicated as it depends on electron momentum and band index in an essential way. However, there is a simplified scaling which applies. One can show that for reduced variables, $\bar{k}=k/k_F$, $\bar{\omega}=\omega/\mu_0$, $\bar{\Sigma}_s(\bar{k},\bar{\omega})$ is a unique function where $\bar{\Sigma}_s\equiv \Sigma_s/\mu_0$. The expression for the self energy has two pieces and is written as
\begin{eqnarray}\label{eq6}
\bar{\Sigma}_s^{EEI}(\bar{k},\bar{\omega})&=&\sum_{s^\prime=\pm 1}\int_{0}^\infty \int_{0}^{2\pi} \frac{d\bar{q}d\theta_{\vec{k}\vec{q}}}{2\pi}\frac{\alpha}{g}
F_{ss^\prime}(\theta_{\vec{k}\vec{k}^\prime}) \nonumber \\&\times&
\Bigg{[}\varepsilon^{-1}(\bar{q},\bar{\omega}-\bar{\epsilon}_{\vec{k}+\vec{q}}^{s^\prime})
\nonumber [\Theta(\bar{\omega}-\bar{\epsilon}_{\vec{k}+\vec{q}}^{s^\prime})-\Theta(-\bar{\epsilon}_{\vec{k}+\vec{q}}^{s^\prime})]
\\
&-&\int_{-\infty}^{\infty}\frac{d\bar{\Omega}}{2\pi}\varepsilon^{-1}(\bar{q}, i\bar{\Omega})\Bigg{(}
\frac{\bar{\omega}-\bar{\epsilon}_{\vec{k}+\vec{q}}^{s^\prime}}{\bar{\Omega}^2+(\bar{\epsilon}_{\vec{k}+\vec{q}}^{s^\prime}-\bar{\omega})^2}-\frac{i \bar{\Omega}}{\bar{\Omega}^2+(\bar{\epsilon}_{\vec{k}+\vec{q}}^{s^\prime}-\bar{\omega})^2}\Bigg{)}\Bigg{]}
\end{eqnarray}
where
\begin{equation}\label{ep7}
F_{ss^\prime}(\theta_{\vec{k}\vec{k}^\prime})
=\frac{1}{2}[1+\cos(\theta_{\vec{k}\vec{k}^\prime})s s^\prime]
\end{equation}
and $\varepsilon^{-1}$ is the inverse dielectric function. It is calculated here in a random phase approximation and takes the form
\begin{equation}\label{eq8}
\varepsilon^{-1}(\bar{q},\bar{\Omega})=\frac{\bar{q}}{\bar{q}-\alpha\bar{\Pi}(\bar{q},\bar{\Omega})}
\end{equation}
with $\bar{\Pi}(\bar{q},\bar{\Omega})$ the polarization function written in reduced variables.\cite{leblanc:2011} The parameter $\alpha = ge^2/(\epsilon_0 v_F)$ gives the overall strength of the coulomb interaction and is inversely proportional to the dielectric function of the substrate $\epsilon_0$ (average of top and bottom medium) and the degeneracy factor $g=4=g_s g_v$.

From Eq.~(\ref{eq1}) and the definitions of $1/\tau^{op}$ and $\lambda^{op}$ we arrive at expressions for these two optical constants which form the basis of our discussion, namely
\begin{eqnarray}
\frac{1}{\tau^{op}(T,\omega)}&=&\frac{\Omega_p^2}{4\pi}{\rm Re}\Bigg{(}\frac{1}{\sigma(T,\omega)}\Bigg{)}\label{eq9}\\
\lambda^{op}(T,\omega)+1&=&-\frac{\Omega_p^2}{4\pi\omega}{\rm Im}\Bigg{(}\frac{1}{\sigma(T,\omega)}\Bigg{)}\label{eq10}.
\end{eqnarray}
We will apply these equations only for the intraband part of the conductivity of Eq.~(\ref{eq3}).  From Eq.~(\ref{eq4}) we see that $A(k,\omega)\equiv A^+(k,\omega)+A^-(k,\omega)$ where $A^+$ describes the upper Dirac cone and $A^-$ the lower cone. The product of $A(k,\omega)A(k,\omega+\Omega)$ then expands into four terms, $A^+(k,\omega)A^+(k,\omega+\Omega)$ and $A^-(k,\omega)A^-(k,\omega+\Omega)$ the intraband terms and $A^+(k,\omega)A^-(k,\omega+\Omega)$ and $A^-(k,\omega)A^+(k,\omega+\Omega)$ are interband terms describing transitions from valence to conduction band. These will be discussed separately later. For now we deal only with the intraband part for which it is natural and useful to define an optical scattering rate and optical effective mass.

\begin{figure}
\vspace*{-1.4cm}%
\centerline{\includegraphics[width=6.0in]{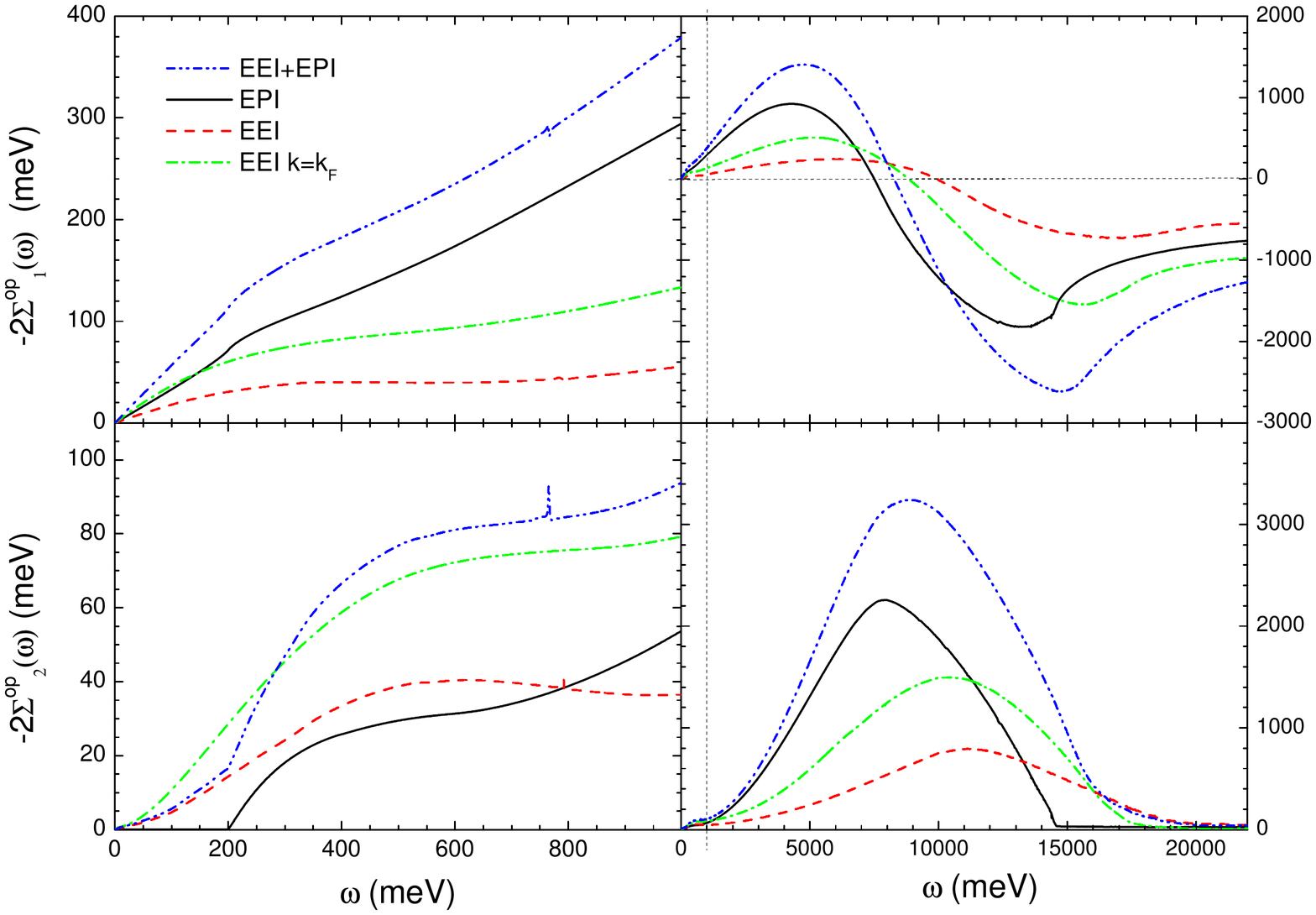}}
\vspace*{-1.0cm}%
\caption{(Color online) The negative of twice the real part $-2\Sigma^{op}_{1}(\omega)$ (top row) and imaginary part $-2\Sigma^{op}_{2}(\omega)$ (bottom row) of the optical self energy calculated from equations (\ref{eq9}) and (\ref{eq10}) for $\sigma(\omega)$ given by the the Kubo formula, Eq.~(\ref{eq3}), for the real part of the conductivity and a Kramers-Kronig transform for the imaginary part. Various cases are considered. A pure electron-phonon case (solid black), a pure electron-electron case (dashed red), a combined EEI+EPI case (dashed double dotted blue) and a simulation of EEI with quasiparticle self energy approximated for all $k$'s by its value pinned at $k=k_F$ (dashed dotted green curve). The left column ranges up to 1~eV while the right column extends up to 22~eV.}
\label{fig1}
\end{figure}

In Fig.~\ref{fig1} we show results for $-2\Sigma^{op}_1(\omega)$ (top row) and $-2\Sigma^{op}_2(\omega)$ (bottom row) both in meV as a function of $\omega$ over an small (left column) and expanded (right column) energy range (1~eV and 22~eV respectively) which extends well beyond the cut off of the bare band. Four distinct cases are shown. Pure EPI case, solid black line, the pure EEI case, dashed red line, and the results when the electronic self energy is fixed in momentum at $k=k_F$ which we denote by $EEI_{k=k_F}$ and is shown as the green dotted-dashed curve and finally, the EEI+EPI case shown as the double-dotted-dashed blue curve. In all cases the $1/\tau^{op}(\omega)$ remains positive definite for all $\omega$ and is interpreted physically to be a scattering rate. By contrast, the real part of the optical self energy ($-2\Sigma^{op}_1(\omega)$) changes signs. In this paper we are mainly interested in the low energy part of this data,  emphasized in the left column, rather than on how interactions modify the high energy band edge. It is important to keep in mind that for bare bands, the optical self energy, $\Sigma^{op}(\omega)$, would be zero. Therefore the deviations of the curves in Fig.~\ref{fig1} from zero carry the information on the effects of correlations. An interesting case to emphasize should be the elastic impurity scattering rate which gives the usual Drude theory. In this case the Kramers-Kronig transformation of a constant scattering gives a zero real part.

\begin{figure}
\vspace*{-1.4cm}%
\centerline{\includegraphics[width=5.0in]{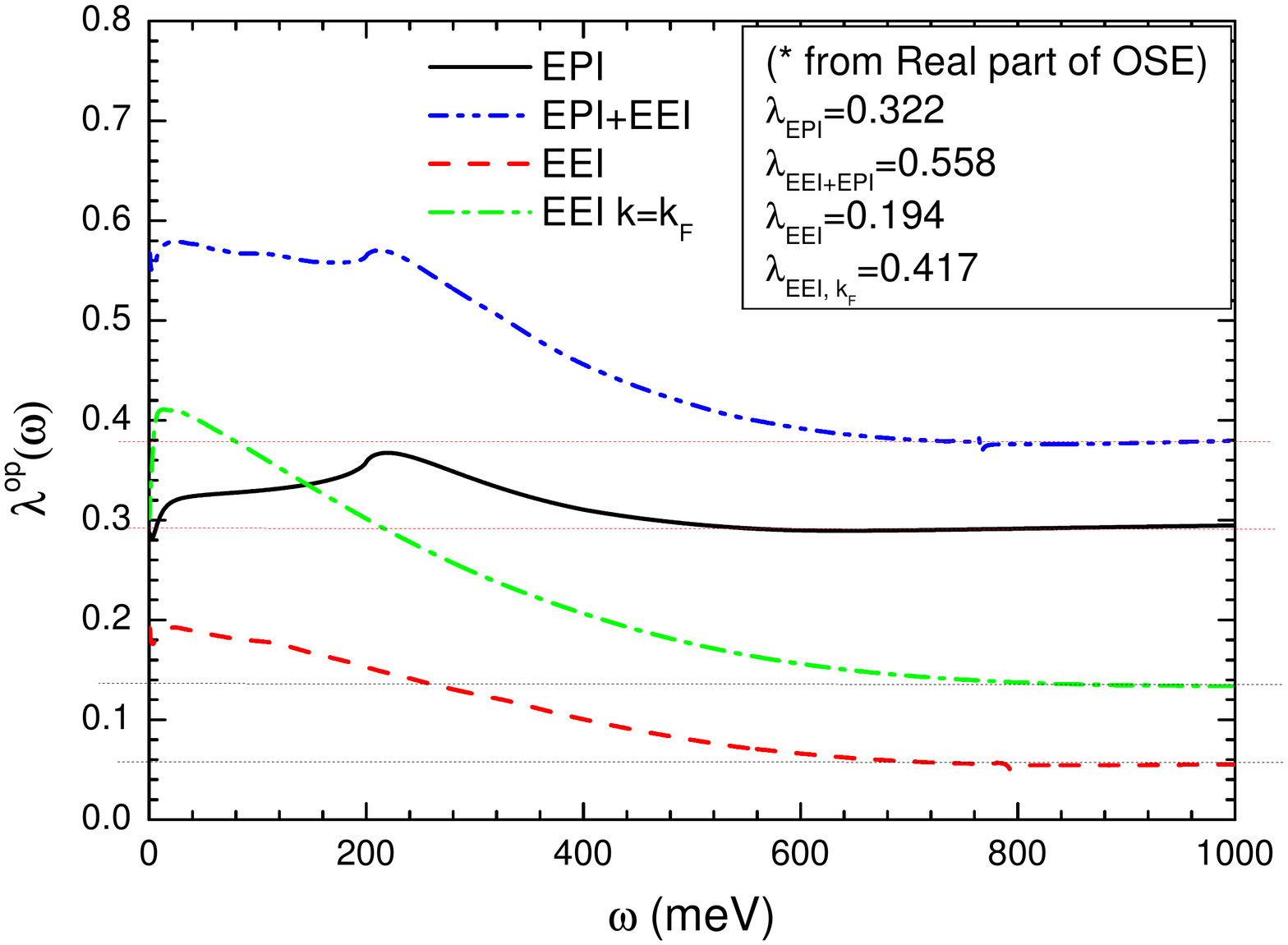}}
\vspace*{-1.0cm}%
\caption{(Color online) The optical effective mass renormalization $\lambda^{op}(\omega)$ vs $\omega$ for pure EPI (solid black), EEI alone (dashed red), combined EPI+EEI (dashed-double dotted blue) and a simulation of EEI alone for which the quasiparticle self energy is approximated for all $k$'s by its value pinned at $k=k_F$.}
\label{fig2}
\end{figure}

In Fig.~\ref{fig2} we show results for the optical effective mass enhancement parameter, $\lambda^{op}(\omega)$, derived from the data of Fig.~\ref{fig1} in the range up to 1.0 eV. We begin our discussion with the EPI only case (solid black). To compute this curve we have used in Eq.~(\ref{eq5}) a delta function at $\omega=\Omega_E$ to represent coupling to a single boson. Note the strong peak feature in $\lambda^{op}(\omega)$ at $\omega=\Omega_E$. This quantity mirrors the sharp structure in the underlying spectral density, $\alpha^2F(\omega)$, and can be used as a first view of the boson exchange spectra involved in the quasiparticle scattering. A rule of thumb is that a peak in the electron-phonon spectral density translates into a corresponding peak in the effective mass renormalization $\lambda^{op}(\omega)$ at this same energy. Also note the rather flat value of $\lambda^{op}$ below $\omega=\Omega_E$ with $\lambda^{op}(\omega)\cong0.32$ to be compared with a value of $\approx 0.29$ at energies above the phonon. This is to be contrasted with the case for the pure EEI, dashed-red, as well as the results of a simulation, dashed-dotted green, for which we ignore the variation of the coulomb self energy with momentum and have instead fixed it to its value at the Fermi momentum, $k=k_F$. In comparison to the black curve, $\lambda^{op}$ shows no peak but rather decreases smoothly from its value near $\omega=0$. This indicates that the effective boson spectrum to which the electrons are coupled is rather unstructured and flat in its energy dependence. Including an additional phonon part to the electron-boson spectral density restores structure to the optical mass renormalization at $\omega=\Omega_E$ as we see in the dash-double-dotted blue curve.

\begin{figure}
\vspace*{-0.50cm}%
\centerline{\includegraphics[width=3.5in]{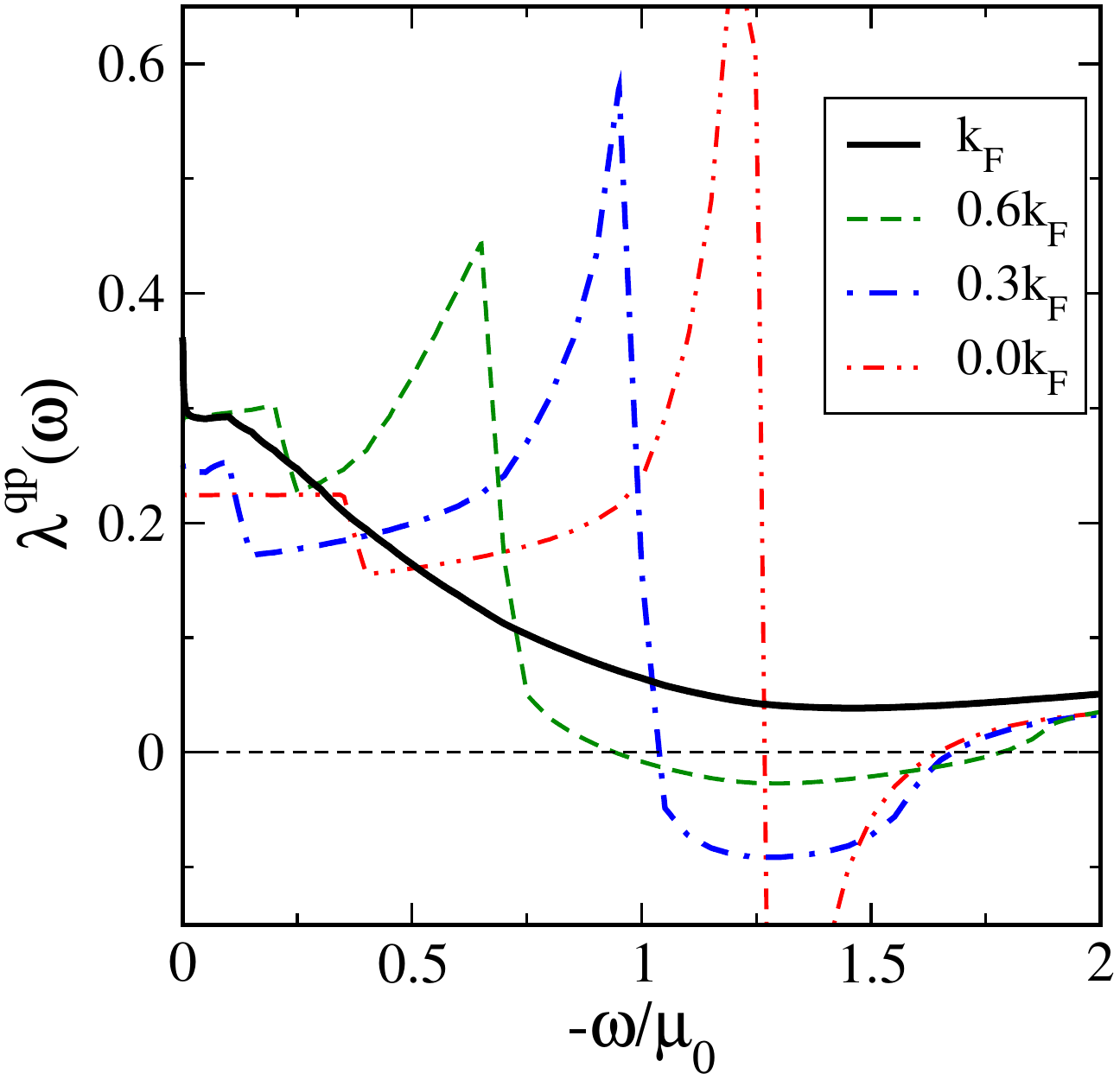}}
\vspace*{-0.0cm}%
\caption{(Color online) The quasiparticle electronic mass renormalization $\lambda^{qp}(k,\omega)$ for four values of momentum as a function of reduced energy $\omega/\mu_0$. The cases are $k=k_F$ (solid black), $k=0.6k_F$ (dashed green), $k=0.3k_F$ (dash-dotted blue) and $k=0$ (dashed double-dotted red).}
\label{fig3}
\end{figure}

While we can make conclusions about the nature of the excitation spectrum involved in the optical renormalization from the resulting shape displayed by $\lambda^{op}(\omega)$ vs $\omega$ we can get additional insight from the study of simulations based on a boson exchange model similar to the electron-phonon model but now for arbitrary choice of $\alpha^2F(\omega)$ chosen to simulate, as well as such a model can, the effects of the EEI. This will be comparable to previous results for marginal Fermi liquid systems. One can also gain insight by comparing our results for the mass renormalization $\lambda^{op}(\omega)$ vs $\omega$ obtained from optics with similar results for the quasiparticle. The quasiparticle self energy, $\Sigma^{qp}(k,\omega)$, depends both on momentum, $k$, as well as on energy, $\omega$, in contrast to its optical analog which depends only on energy since the conductivity of Eq.~(\ref{eq3}) involves an integration over $\vec{k}$. Results for the quasiparticle effective mass $\lambda^{qp}(k, \omega)\equiv\frac{1}{\omega}[\Sigma^{qp}(k,\omega)-\Sigma^{qp}(k, \omega=0)]$ are shown in Fig. \ref{fig3} for four values of $k$ namely $k=k_F$ (solid black line), $k=0.6k_F$ (dashed green line), $k=0.3k_F$ (dashed-dotted blue) and $k=0.0$ (doubled dotted dashed red line). Note that the overall behavior of $\lambda^{qp}(k,\omega)$ for $k=k_F$ is very similar to that found in the pervious Fig. \ref{fig2} dashed dotted green curve for the optical effective mass renormalization $\lambda^{op}(\omega)$, which applies for a value of chemical potential equal to 400 meV. While the magnitude of the corresponding self energy is larger by about 1/3, its variation with $\omega$ is much the same. More importantly as we move away from $k=k_F$ towards zero momentum, the corresponding structure in the quasiparticle curves does not look like the optical quantity. This is a first indication that optics (intraband) is most sensitive to the boson structure or excitation spectrum which dresses the quasiparticle at the Fermi surface, $k=k_F$, rather than at the Dirac point near $k=0$. This idea is investigated in the next section through maximum entropy inversion techniques.

\section{Approximate Boson Exchange Model}

To help us understand better the many-body interactions at play in graphene we have found it useful to introduce an approximate but analytic boson exchange model which was previously derived and applies for any energy dependent density of state $N(\omega)$. Here we specialize it to the specific case of graphene for which $N(\omega)/N_0=|\omega+\mu_0|/\mu_0$.  The  formulas for quasiparticle self energy given by Eq.~(\ref{eq5}) can be worked out to read
\begin{eqnarray}\label{eq11}
Re\Sigma^{qp}(\omega)&=&\int_{0}^{\infty}d \Omega \frac{\alpha^2F(\Omega)}{\mu_0} \Big{\{}-2\mu_0+(\omega+\mu_0)\ln\Big{|}
\frac{(\Omega+\omega+\mu_0)^2(\Omega-\omega)}{(\Omega+\omega)(\omega+\Omega+W_c)(W_c+\Omega-\omega)}\Big{|}  \nonumber \\
&+&\Omega \ln \Big{|}\frac{(\Omega+\omega+\mu_0)^2}{(\Omega+\omega)(\Omega+\omega+W_c)}\frac{(W_c+\Omega-\omega)}{(\Omega-\omega)} \Big{|}  \Big{\}}.
\end{eqnarray}
For the quasiparticle mass renormalization parameter, $\lambda^{qp}(\omega)$, given by
\begin{equation}\label{eq12}
-\omega \lambda^{qp}(\omega)\equiv {\rm Re}\Sigma^{qp}(\omega)-{\rm Re}\Sigma^{qp}(0)
\end{equation}
we evaluate the case of $\omega =0$ to obtain
\begin{equation}\label{eq13}
\lambda^{qp}(\omega=0)=2\int^{\infty}_{0} d \Omega\alpha^2F(\Omega) \Big{\{} \frac{1}{\Omega}-\frac{1}{\mu_0}
+\frac{\Omega}{\mu_0(\Omega+W_c)}+\frac{1}{\mu_0}\ln\Big{|} \frac{W_c+\Omega}{\mu_0+\Omega} \Big{|}\Big{\}}.
\end{equation}

We emphasize that the first term gives the spectral effective mass $\lambda\equiv 2\int^{\infty}_{0}\alpha^2F(\Omega)/\Omega d\Omega$ which is the only quantity that would enter in a conventional metal with constant density of states.  In general the quasiparticle mass renormalization, $\lambda^{qp}(\omega=0)$, differs from its spectral value, $\lambda^{op}(\omega=0)$. The imaginary part of the quasiparticle self energy of Eq.~(\ref{eq5}) which gives the quasiparticle scattering rate can also be simplified. It reads
\begin{equation}\label{eq14}
-{\rm Im} \Sigma^{qp}(\omega)=\pi \int^{\omega}_{0} d\Omega \alpha^2F(\Omega)N(\omega-\Omega)
\end{equation}
for $\omega > 0$ and
\begin{equation}\label{eq15}
-{\rm Im} \Sigma^{qp}(\omega)=\pi \int^{|\omega|}_{0} d\Omega \alpha^2F(\Omega)N(-|\omega|+\Omega)
\end{equation}
for $\omega < 0$. Note that here we use the density of states of doped graphene with a linear band approximation as $N(\omega)=|\omega+\mu_0|/\mu_0$. The resulting quasiparticle scattering rate is not necessarily the same for negative and positive $\omega$. In the limit of $\omega \rightarrow 0$ we assume that the boson spectral density contains a constant $\omega^0$ term, $\alpha^2F(\omega) = C + O(\omega)$, and neglect linear and higher order powers of $\omega$. With this we obtain
\begin{equation}\label{eq16}
-{\rm Im} \Sigma^{qp}(\omega) = \pi C |\omega|
\end{equation}
which is linear in $\omega$ for constant $C$.

Similar formulas have been derived for the optical mass renormalization and scattering rate which when applied to the specific case of graphene give
\begin{equation}\label{eq17}
\lambda^{op}(\omega)=\frac{2}{\omega^2}\int^{\infty}_{0}d\Omega \alpha^2F(\Omega)P\int^{\infty}_{0}d\omega' \frac{|\omega'+\mu_0|}{\mu_0}
\ln\Big{|} \frac{(\omega'+\Omega)^2}{(\omega'+\Omega)^2-\omega^2}\Big{|}
\end{equation}
and in the limit $\omega \rightarrow 0$ we get back the same result as Eq.~(\ref{eq13}) namely $\lambda^{op}(\omega=0)\equiv\lambda^{qp}(\omega=0)$. One should note that for finite $\omega$, Eq.~(\ref{eq17}) can be solved as a series of logarithms. The simplified approximate formulas derived above can be used to analyze the numerical results obtained from the evaluation of the full Kubo formula for the conductivity described in the previous section and to obtain insight into the nature of the effective excitation spectrum that scatters the electrons when electron-electron interactions are accounted for. The expression for the optical scattering rate is
\begin{eqnarray}\label{eq18}
\frac{1}{\tau^{op}(\omega)}&=&2\pi \int^{\omega}_{0} \alpha^2F(\Omega)\Big{(}\frac{\omega-\Omega}{\omega}\Big{)} \:\:\:\:\:\:\mbox{for} \:\:
\Omega<\omega <\mu_0+\Omega \nonumber \\
&=&\frac{\pi \mu_0}{\omega} \int^{\omega}_{0} \alpha^2F(\Omega)\Big{[}1+\Big{(}\frac{\omega-\Omega}{\mu_0}\Big{)}^2\Big{]} \:\:\:\mbox{for} \:\:
\omega>\mu_0+\Omega.
\end{eqnarray}
The zero $\omega$ limit of $1/\tau^{op}(\omega)$ for the same model $\alpha^2F(\Omega)$ as previously used gives $1/\tau^{op}(\omega)=\pi C|\omega|$ which is the same as its quasiparticle counterpart.

\section{Recovering a Boson Spectrum}

\begin{figure}
\vspace*{-1.4cm}%
\centerline{\includegraphics[width=5.0in]{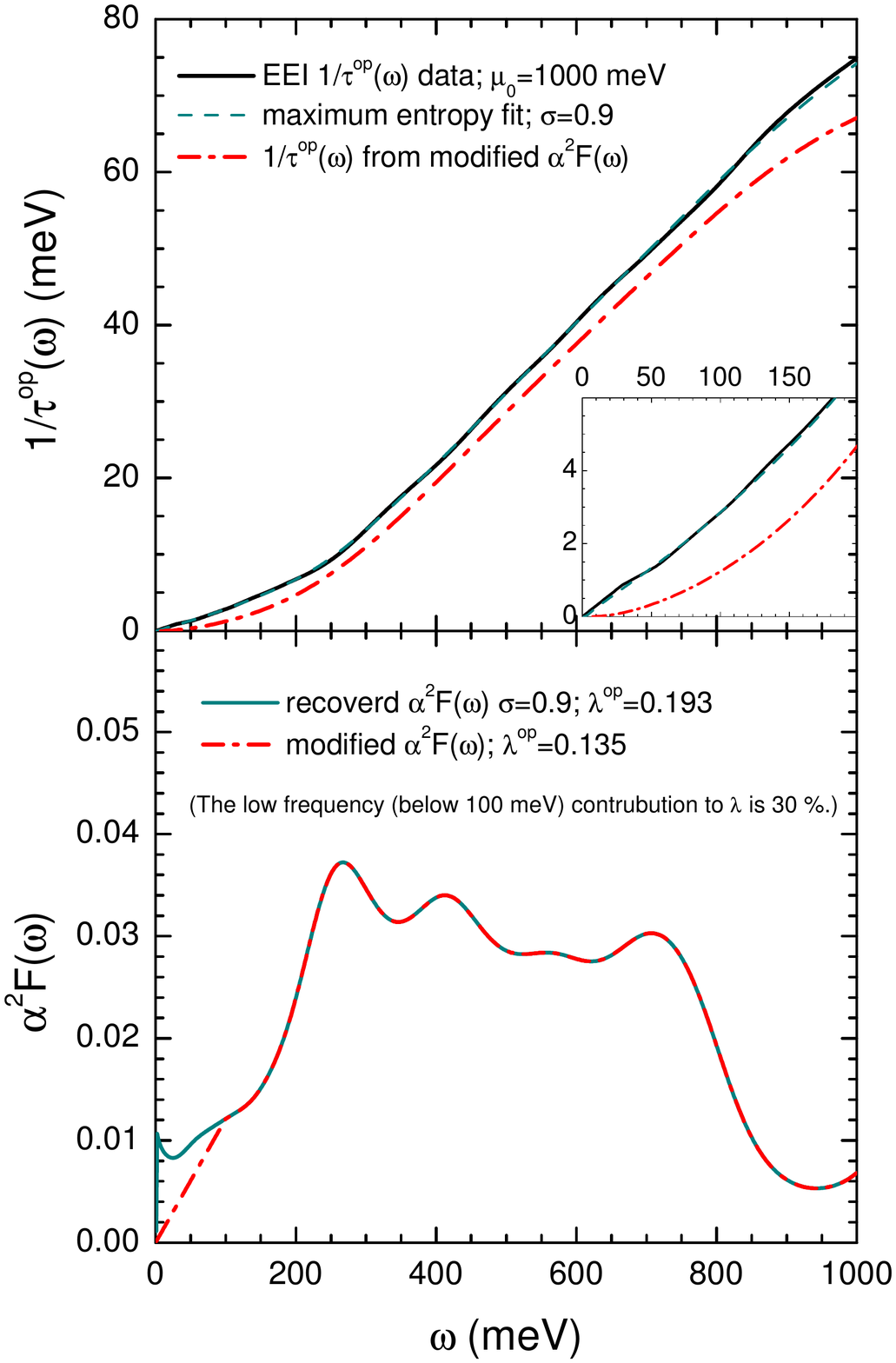}}
\vspace*{-0.5cm}%
\caption{(Color online) The top frame displays the optical scattering rate $1/\tau^{op}(\omega)$ (in meV) as a function of $\omega$ (in meV) for the case of pure EEI (solid black).  The bottom frame displays the various  $\alpha^2F(\omega)$ obtained through maximum entropy inversion which are used to obtain fits to the input EEI of the top frame.   The dashed blue curve is the maximum entropy fit to the solid curve while the dash-doted red curve (which does not fit to the data well) is obtained by removing spectral density at low $\omega$, described in detail in the text, and shown in the lower frame as the dashed-dotted red curve assumed linear in $\omega$ below 100 meV. The solid green curve is the electron boson spectral density obtained from our maximum entropy fit. The inset in the top frame provides an expanded view of the scattering rates at small $\omega$.}
\label{fig4}
\end{figure}

\begin{figure}
\vspace*{-1.4cm}%
\centerline{\includegraphics[width=5.0in]{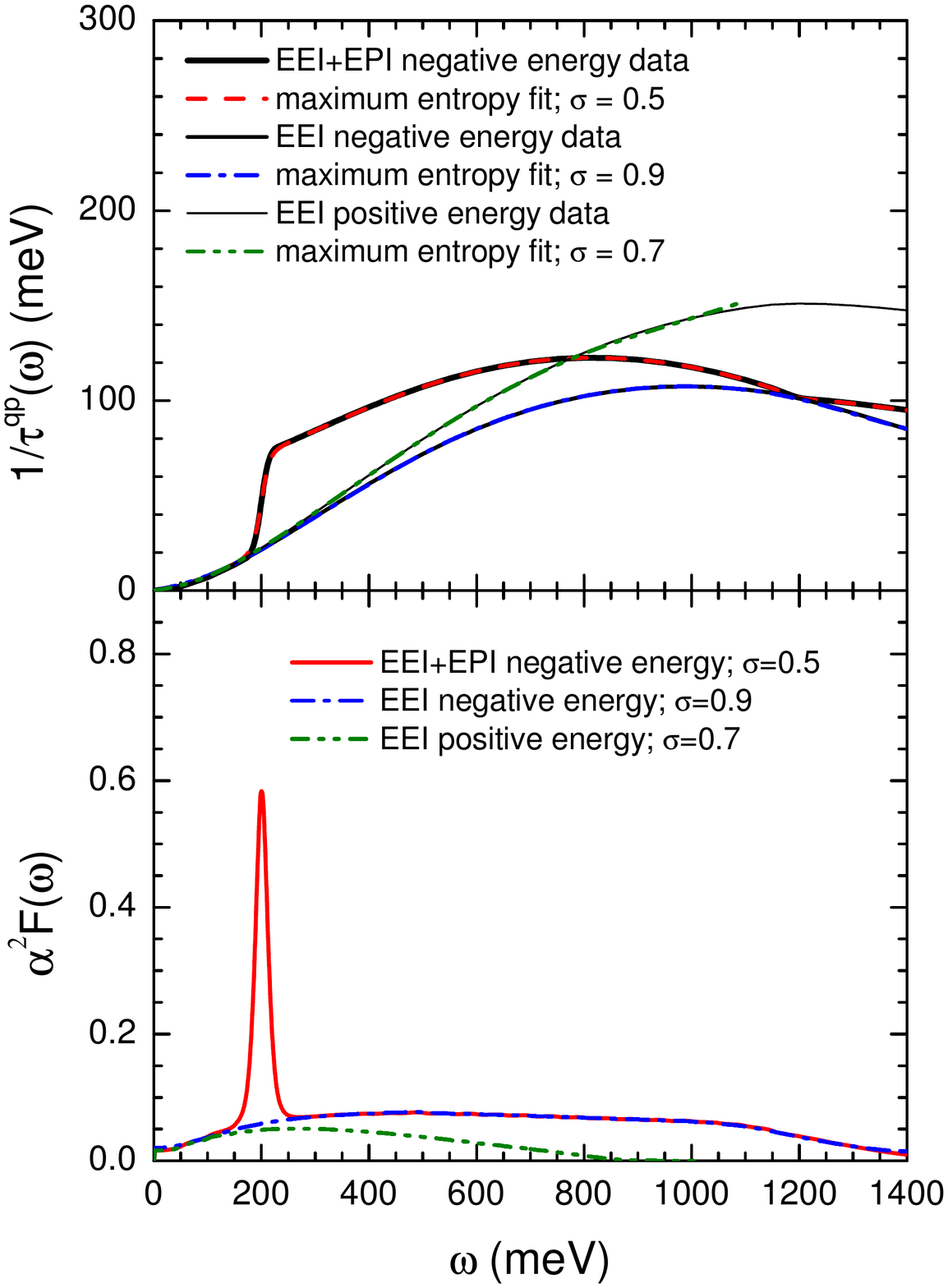}}
\vspace*{-1.5cm}%
\caption{(Color online) Upper frame, the quasiparticle scattering rate $1/\tau^{qp}(\omega)$ (in meV) vs $\omega$ (in meV) for various cases.  Negative and positive data refers to sign of frequency $\omega$ as the quasiparticle scattering rate can be different for positive and negative frequency cases.
The dashed-dotted blue curve is the inversion fit to the negative energy EEI data; the dashed double-dotted  green curve is the fit to the positive energy EEI data; and the red dashed curve is the inversion fit to the EEI+EPI case for negative energies. Lower frame, the electron boson spectral densities recovered from the maximum entropy fit for EEI+EPI (solid red), EEI negative frequencies (dashed dotted blue) and positive frequencies (dashed double-dotted green).}
\label{fig5}
\end{figure}

\begin{figure}
\vspace*{-1.4cm}%
\centerline{\includegraphics[width=5.0in]{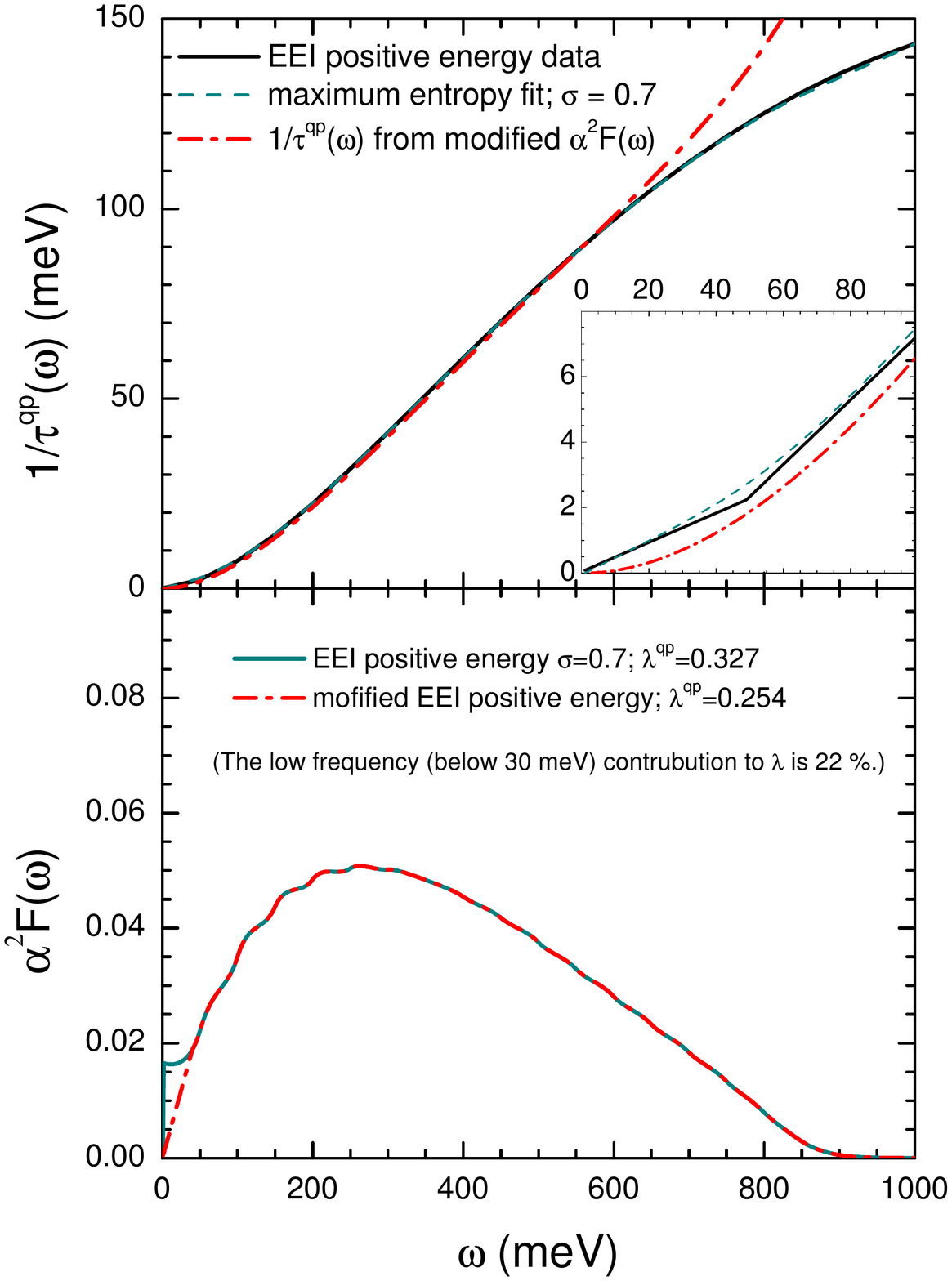}}
\vspace*{-1.5cm}%
\caption{(Color online) Top frame, the quasiparticle scattering rate $1/\tau^{qp}(\omega)$ (in meV) vs $\omega$ (in meV) for EEI alone (solid black), its maximum entropy fit (dashed blue) and the results of a modification in the electron boson spectral density (dash-dotted red). The inset provides an expanded view of the low $\omega$ region. Lower frame, the recovered electron boson spectral density solid blue curve through maximum entropy inversion of the data of the top frame. The dash-dotted red curve is the same but with an arbitrary modification {\it i.e.} linear dependence in $\omega$ from 0 to 50 meV.}
\label{fig6}
\end{figure}

It is useful to use equation (\ref{eq18}) in a maximum entropy inversion of our numerical data for the optical scattering rate which includes the electron-electron interaction shown in Fig. \ref{fig1}. For a general kernel, $K(\omega,\Omega)$, and input data, $I(\omega)$, with $I(\omega)=\int^{+\infty}_{-\infty}K(\omega,\Omega)\alpha^2F(\Omega)d\Omega$ the deconvolution of this equation to recover an effective spectral density, $\alpha^{2}F(\Omega)$ is ill conditioned and here we use a maximum entropy technique\cite{schachinger:2006}. We begin by discretizing the equation to $I(i) = \sum_j K(i,j)\alpha^{2}F(j)\Delta \Omega$ where $\Delta \Omega$ is the differential increment on the integration over $\Omega_j=j\Delta \Omega$. We define a $\chi^2$ by
\begin{equation}\label{eq19}
\chi^2=\sum_{i=1}^{N}\frac{[I(i)-\Sigma(i)]^2}{\sigma_i^2}
\end{equation}
where $I(i)$ is the input data, $\Sigma(i) \equiv \sum_j K(i,j) \alpha^2F(j)$ the calculated value from the known kernel and a given choice of $\alpha^{2}F(\Omega)$, and $\sigma_i$ is the error assigned to the data $I(i)$. Appropriate constraints such as positive definiteness for the boson exchange function, $\alpha^{2}F(\Omega)$, are introduced and the entropy functional
\begin{equation}\label{eq20}
L=\frac{\chi^2}{2}-aS
\end{equation}
where the Shannon-Jones entropy, $S$,\cite{schachinger:2006} is minimized with
\begin{equation}\label{eq21}
S=\int^{\infty}_{0}\Big{[} \alpha^{2}F(\Omega)-m(\Omega)-\alpha^{2}F(\Omega)\ln\Big{|}\frac{\alpha^{2}F(\Omega)}{m(\Omega)} \Big{|}\Big{]}.
\end{equation}
The parameter $a$ in equation (\ref{eq20}) controls how close a fit to the data is obtained. The parameter $m(\Omega)$ is here taken to be some constant value on the assumption that there is no a priori knowledge of the functional form of the electron-phonon spectral density $\alpha^2F(\Omega)$. While there is no guarantee that a boson exchange model can successfully reproduce consistently, quantitatively, and accurately all the details of our calculated data for the optical self energy in the case of the EEI application of the inversion process can still provide useful insight into the excitation spectrum responsible for the scattering in this case.

In Fig. \ref{fig4} we show results obtained from the optical scattering rate $1/\tau^{op}(\omega)$ of Eq.~(\ref{eq9}) which is based on the Kubo formula, Eq.~(\ref{eq3}), and the electron spectral density, Eq.~(\ref{eq4}), with quasiparticle self energy given by equation (\ref{eq6}) which is based on a random phase approximation for the dynamical screening in graphene. A Kramers-Kronig transform is also implied. This is how the imaginary part of the conductivity is obtained from its real part given in equation (\ref{eq3}). Both real and imaginary parts of the optical conductivity are required in constructing the optical scattering rate. In the top frame the input data on $1/\tau^{op}(\omega)$ in meV as a function of $\omega$ up to 1000 meV is shown (heavy solid black). The maximum entropy fit for $\sigma = 0.9$ is shown (dashed blue) and is seen to be an excellent fit over the entire range of $\omega$ considered. Here the chemical potential was set at $\mu_0=$ 1000 meV for definiteness. The recovered spectral density $\alpha^{2}F(\omega)$ is shown in the lower frame (solid blue). An important feature to notice is that below 100 meV $\alpha^{2}F(\omega)$ remains finite to rather low energies before going to zero at very low $\omega$ (not seen on the scale used for this figure). This is characteristic of the Coulomb interaction and reminds one of the marginal Fermi liquid model for which $\alpha^{2}F(\omega)$ is constant for all $\omega$ greater than a low energy cutoff on the order of the temperature. It is interesting to look, at the low $\omega$ behavior more closely. The dashed dotted red curve shows a modified spectral density which has been made to vanish linearly below 100 meV. For this spectrum we obtain the dashed-dotted red curve for the scattering rate shown in the top frame. We see that this arbitrary modification has significantly altered our low energy fit and that to get agreement with the input data a reasonably constant value of the spectral density is indeed needed {\it i.e.} a marginal Fermi liquid like behavior is indicated. A reasonable first measure of the difference at small $\omega$ in the two spectra of Fig.~\ref{fig4} (lower frame) is the value of the spectral lambda, $\lambda \equiv 2\int^{\infty}_0  \frac{\alpha^2F(\omega)}{\omega} d\omega $, which emphasizes this region. In the best fit case $\lambda = 0.193$ while in the other it is reduced to 0.135 but clearly the first of these values is preferred as the low $\omega$ value of the data for $1/\tau^{op}(\omega)$ shows linear in $\omega$ dependence in the range shown and this is taken as characteristic of coulomb interactions, as emphasized in the inset which shows a magnified view of the low $\omega$ dependence only.

A scattering rate taken to be linear in $\omega$ at small $\omega$ implies a constant value of $\alpha^2F(\omega)$ at small $\omega$.  This is true for both optical and quasiparticle quantities. In Fig. \ref{fig5} we show results for the imaginary part of the quasiparticle self energy. As noted in equations (\ref{eq14}) and (\ref{eq15}) the form of the integral determining this quantity in graphene depends on the sign of $\omega$. For finite chemical potential $\mu_0$, even the density of electronic states is not symmetric about $\omega=0$. In the top frame of Fig. \ref{fig5} we show results for $1/\tau^{qp}(\omega)$  for both positive and negative $\omega$ data for EEI alone and for a combined case of EEI+EPI for the negative range only (heavy solid black curves). In all cases the maximum entropy fits (shown in the top frame) are very good. These are calculated from the recovered electron-boson spectral densities presented in the lower frame. We note first that all curves remain finite at small $\omega$ on the scale of the figure. Of course the maximum entropy inversion respects the constraint that $\alpha^{2}F(\omega) = 0$ at $\omega = 0$. As $\omega$ is increased the two pure EEI spectra show a broad maximum before dropping towards zero above $\omega \sim$ 1400 meV for the negative energy and around 800 meV for the positive case. The chemical potential is 1000 meV. The third spectrum, shown as the solid red curve, has the same EEI background as the pure electron case but has an additional sharp peak due to coupling to an Einstein phonon at $\omega=\Omega_E= 200$ meV. This EPI part is clearly seen over the EEI background. It is this EPI part which is responsible for the sharp rise in the $1/\tau^{qp}(\omega)$ curve of the top frame for this case. Should the coupling be to a distribution of phonons rather to an Einstein mode then the rise at $\omega=\Omega_E$ would be more gradual, reflecting the details of the distributed phonon spectrum involved. It is clear, on comparison of Fig.~\ref{fig6} for quasiparticle self energies with the data in Fig.~\ref{fig4} for the optical case, that a boson exchange theory does not provide a perfectly consistent picture between these two data sets. Nevertheless the shape of the recovered spectra for EEI are quite similar and neither shows any evidence for important coupling to plasmon structure and both are more characteristic of coupling to a particle-hole continuum. In this regard it is important to note that the quasiparticle data inverted in Fig.~\ref{fig6} is for positive energy data at $k=k_F$. In the upper frame of Fig.~\ref{fig6} we repeat our maximum entropy fit to the scattering rate data for positive energy but also add a second (red) dashed-dotted curve for a case where the $\alpha^{2}F(\omega)$ has been altered at small $\omega$ to go linearly to zero below 50 meV rather than stay constant for the maximum entropy inverted curve (solid blue). This modification of the spectrum has changed the spectral $\lambda$ from 0.327 to 0.254. We note that these two values of mass enhancement recovered from quasiparticle data, while of the same order of magnitude as these recovered from optics, are definitely larger than their optical counterparts. Thus a single electron-boson spectrum does not fit simultaneously both quantities. It is important to recognize, however,  that the shape of the functions involved are not different. We are mainly noting a magnitude difference. In the inset of the top frame of Fig. \ref{fig6} we show the deterioration in our fit to $1/\tau^{qp}(\omega)$ data that our arbitrary modification of the low $\omega$ behavior of the spectral density has caused, i.e. when the red dash-dotted curve for the modified $\alpha^2F(\omega)$ is made linear below 50 meV with no other changes. We emphasize that once again the quasiparticle scattering rate shows a linear in $\omega$ dependence at small energies, reminiscent of the marginal fermi liquid model characteristic of coulomb interactions in this case.

\begin{figure}
\vspace*{-0.0cm}%
\centerline{\includegraphics[width=5.0in]{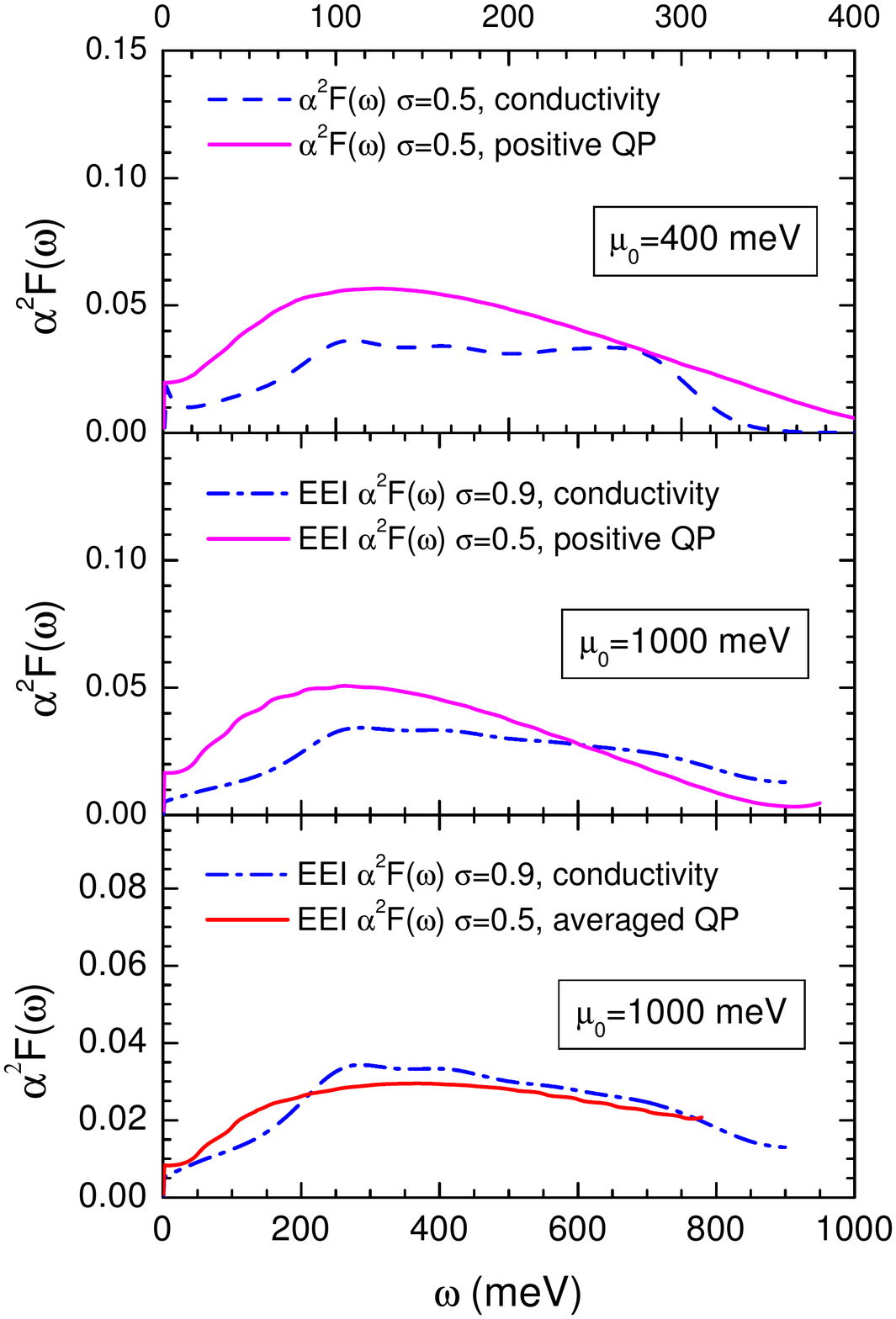}}
\vspace*{-1.0cm}%
\caption{(Color online) Top frame, the maximum entropy results for the electron boson spectral density, $\alpha^2F(\omega)$, obtained from positive energy quasiparticle data (solid red) compared with that obtained from optical data (dashed blue) for chemical potential $\mu_0=$~400~meV. The middle frame is the same but for the case $\mu_0=$ 1000 meV. The lower frame compares optical and quasiparticle spectra for $\mu_0=$~1000~meV with value of mass enhancement $\lambda$ for the quasiparticle case adjusted to be the same as for the optics. Here it is the average of positive and negative energy quasiparticle inversions that is employed.}
\label{fig7}
\end{figure}

In Fig. \ref{fig7} we compare our results for the electron-boson spectral density obtained from positive $\omega$ quasiparticle self energy (solid red) with our optical conductivity data (dashed blue), for two values of chemical potentials. The top frame has $\mu_0=$ 400 meV and the middle flame is for $\mu_0=$ 1000 meV. Comparing these two frames we conclude that in both the quasiparticle and optical cases the boson spectral density does not change its magnitude when the value of $\mu_0$ is changed and that the horizontal energy scale $\omega$ is itself proportional to $\mu_0$. Also, as we have previously noted, the shapes of the extracted $\alpha^{2}F(\omega)$ are not different, rather there is only a scale difference between quasiparticle and optics. In the lower frame of Fig. \ref{fig7} we make this point more clearly. There the dash-dotted blue and solid red are respectively the optical spectral density and the averaged positive and negative energy quasiparticle spectral density with this latter quantity scaled to result in the same spectral $\lambda$ value. The curves are not appreciably different and indicate that in both quantities we are coupled to the same excitations.

\begin{figure}
\vspace*{-0.50cm}%
\centerline{\includegraphics[width=5.0in]{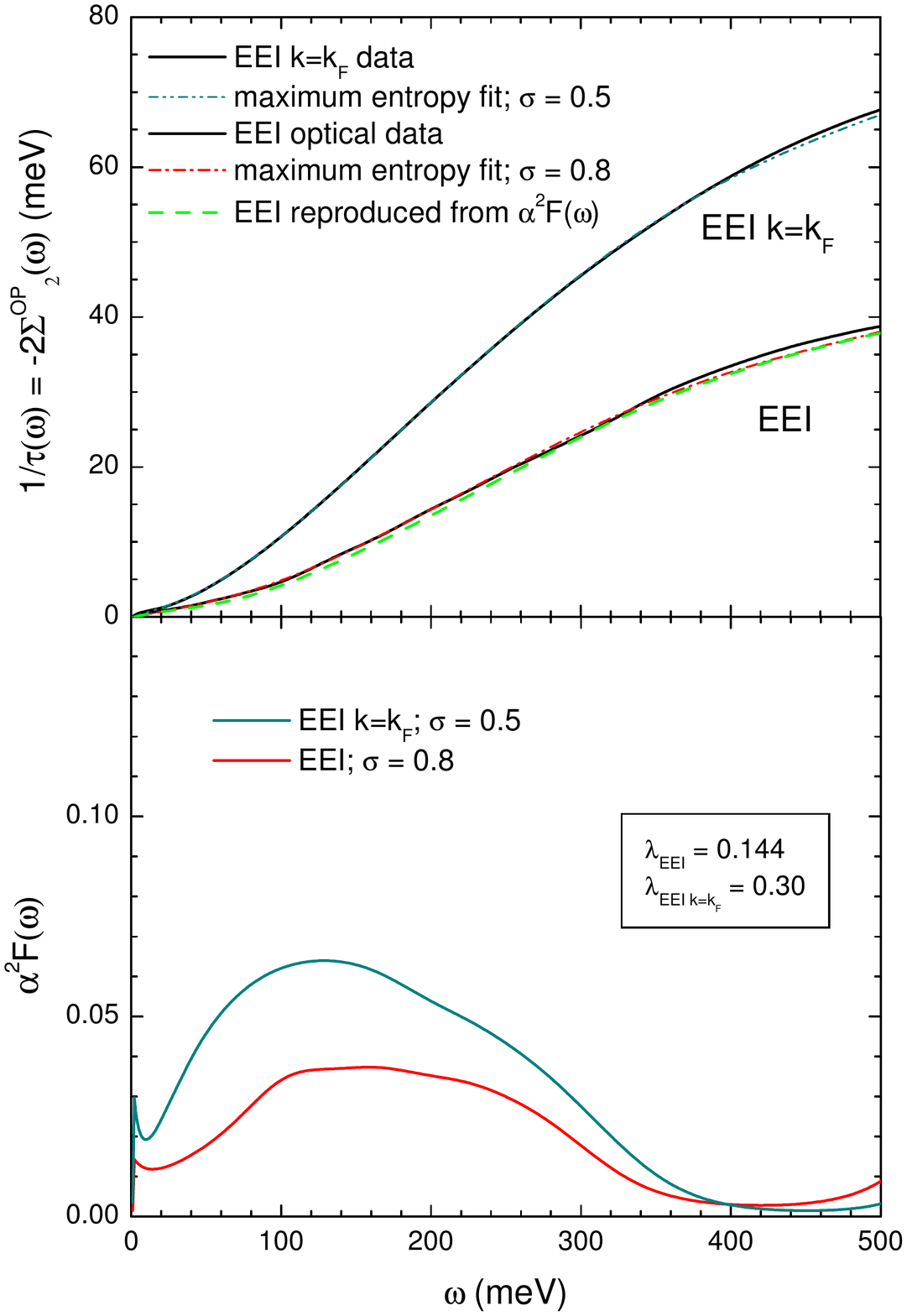}}
\vspace*{-1.5cm}%
\caption{(Color online) Top frame shows the optical scattering rate $1/\tau^{op}(\omega)$ (in meV) vs $\omega$ (in meV) for EEI (solid black) with maximum entropy fit (dash-dotted red).  The lower frame shows the recovered electron boson spectral density, $\alpha^2F(\omega)$, for pure EEI (solid red) and the pinned $k=k_F$ case (solid blue). The solid blue is data for EEI but with the electron-electron self energy pinned at $k=k_F$ for all values of $k$ and the fit is the dashed double dotted blue curve. }
\label{fig8}
\end{figure}

For Coulomb correlations the quasiparticle self energy depends on magnitude of the momentum. In Fig. \ref{fig8} we show results for the optical scattering rate when this momentum dependence is pinned to its value at $k=k_F$ for all $k$'s in Eq.~(\ref{eq6}). As is clear from the top frame this approximation strongly influences the absolute magnitude of the scattering. Comparing the pinned (EEI, $k=k_F$) and regular (EEI) cases we see that the curves do not differ much when one considers their dependence on $\omega$. Both can be fit to a boson exchange theory using maximum entropy methods (dashed double-dotted blue curve for EEI $k=k_F$ and dashed dotted red for EEI). Results for the recovered spectral density $\alpha^{2}F(\omega)$ are shown in the lower frame. The upper blue curve with $\lambda_{EEI, k_F}=$ 0.30 has very much the same shape as the lower red curve for EEI with $\lambda_{EEI}=$ 0.144, although their $\lambda$'s differ by a factor of 2. We interpret this as strong evidence that in optics the boson spectrum of excitations that is sampled is that due to the electron at $k=k_F$. By implication the plasmaron structures corresponding to $k$ near zero, are not important as we have already conclude from consideration of the frequency dependence of the optical effective mass. Extracting the entire spectrum of excitations through inversion has allowed us to address this issue in more detail than has previously been possible.

\begin{figure}
\vspace*{-0.0cm}%
\centerline{\includegraphics[width=5.0in]{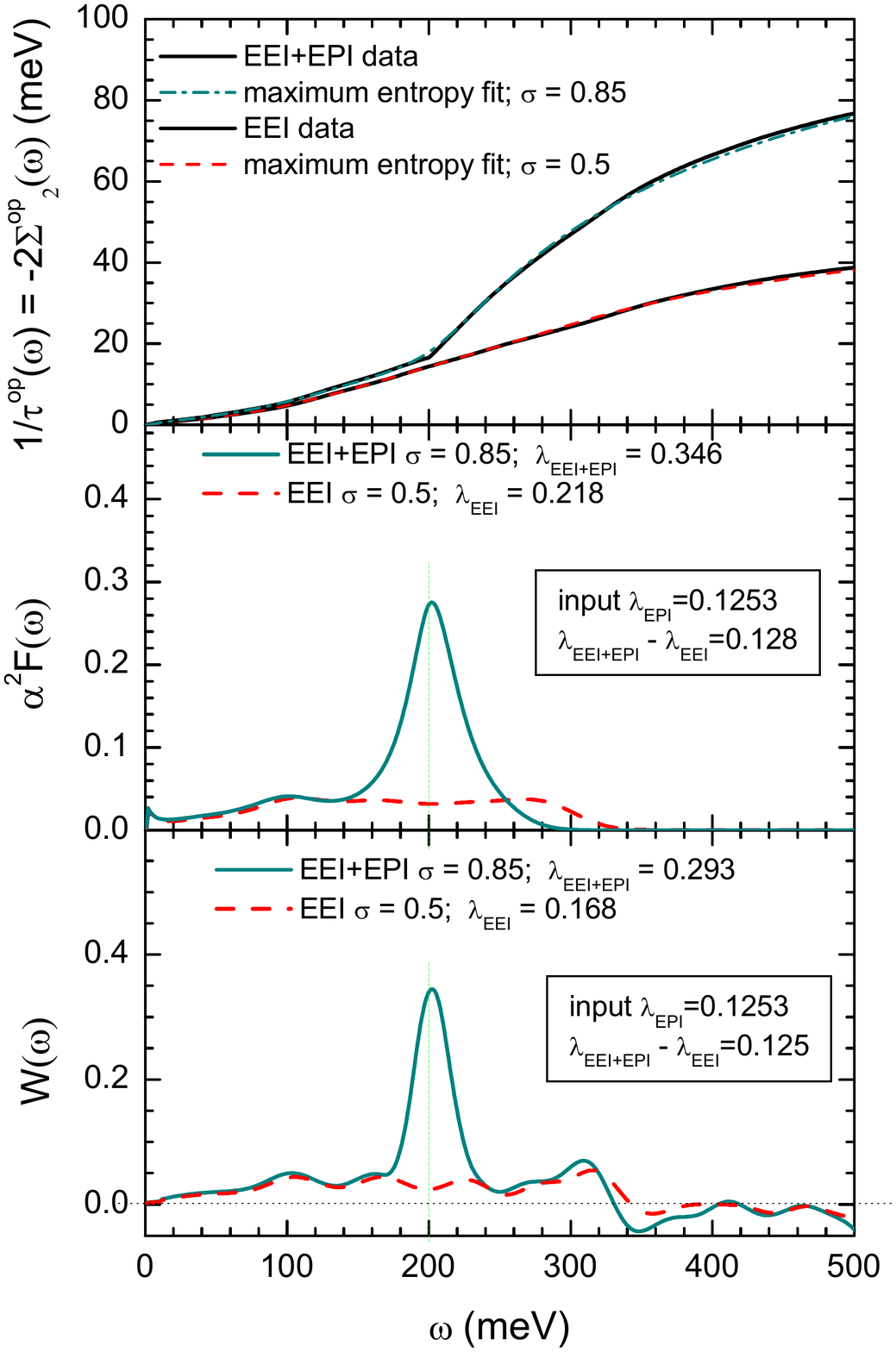}}
\vspace*{-1.0cm}%
\caption{(Color online) Top frame shows the optical scattering rate $1/\tau^{op}(\omega)$ (in meV) vs $\omega$ (in meV) for the case of combined EEI+EPI (solid black) with the maximum entropy fit to the numerical data (dash-dotted blue). This is to be compared with the pure EEI case (solid black) with its fit (red dashed curve). Middle frame gives the electron boson spectral densities recovered from the maximum entropy inversions. The solid blue is EEI+EPI while the dashed red is for EEI alone. This last curve provides a background above which there is an additional phonon peak at 200 meV for the combined case. The bottom frame is the same as middle frame but now showing $W(\omega)$, the second derivative of $\frac{1}{2\pi}\frac{\omega}{\tau^{op}(\omega)}$ with respect to $\omega$, which is used to get an estimate of the underlying spectral density in the combined EEI+EPI (solid blue) and EEI (dashed red) cases. }
\label{fig9}
\end{figure}

Another important point is made in Fig. \ref{fig9}. In the top frame we show results for the optical scattering rate $1/\tau^{op}(\omega)$ in meV as a function of $\omega$ to 500 meV. The low solid black line is for EEI alone which the upper black line includes an electron phonon component (EEI+EPI). Note the sharp rise in $1/\tau^{op}(\omega)$ at $\omega$ = 200 meV which is the energy at which we have placed the Einstein phonon. The dashed red curve is our maximum entropy fit to the lower black curve and the dash-dotted blue to the upper one. In both curves the fits are very good over the entire energy range considered. The recovered electron-boson spectral densities are shown in the middle frame. The dashed red curve is for EEI alone and the solid blue includes both EEI and EPI. Note that the inversion procedure provides almost exactly the input value used for the mass renormalization coming from the phonons in that $\lambda_{EEI+EPI}-\lambda_{EEI}=0.128$ to be compared with the input value $\lambda_{EPI}=0.125$. While the Einstein contribution is characteristically peaked about $\omega=\Omega_E$ = 200 meV, the EEI provides a corresponding smooth and low amplitude background extending to $\omega \geq 300$ meV on which the phonon peak sits. In the lower frame we show additional results for the recovered electron-boson spectral density in the same case but obtained from a second derivative technique \cite{marsiglio:1998,marsiglio:1999} rather than by maximum entropy inversion \cite{schachinger:2006}. It is well known that in conventional metals in which electrons are coupled by the electron phonon interaction the second derivative of the scattering rate times $\omega$ namely $\frac{1}{2\pi}\frac{d^2}{d\omega^2}\Big{(}\frac{\omega}{\tau^{op}(\omega)}\Big{)}\equiv W(\omega)$ is closely related to the input $\alpha^2F(\omega)$ in the energy range where it is non zero. Above the cutoff in $\alpha^2F(\omega)$, $W(\omega)$ shows additional negative tails not part of the original spectrum. These results however depend on the assumption that the electronic density of states does not vary significantly on the energy scale involved in $\alpha^2F(\omega)$. Nevertheless as seen in the lower frame of the figure this also holds reasonably well for graphene. Below roughly 320 meV, solid blue and dashed red curves for $W(\omega)$ are qualitatively very similar to the solid blue and dashed red curves for $\alpha^2F(\omega)$ shown in the middle frame. Both techniques reproduce well the main features of the underlying spectrum. Also, separate signature of both phonons and EEI are clearly seen although strictly speaking the spectra are superimposed. But, because they have quite distinct energy variations this allows for a reasonable separation of the two. There are quantitative differences such as the cutoff around 300 meV occurring at lower energy in the maximum entropy inversion than in the second derivative technique.

\section{Boson Structure in Interband Conductivity}

\begin{figure}
\vspace*{-0.0cm}%
\centerline{\includegraphics[width=3.0in]{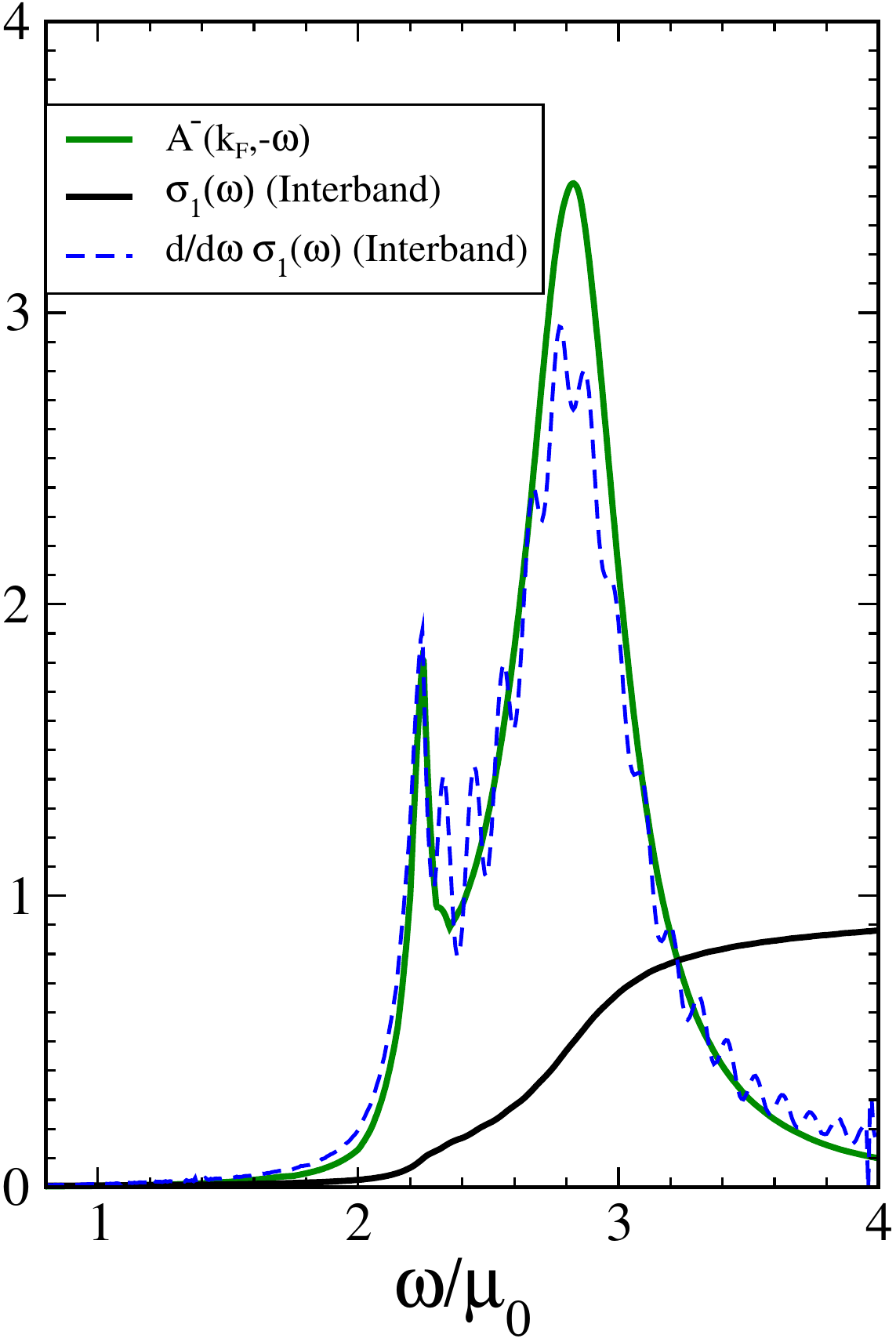}}
\vspace*{-0.0cm}%
\caption{(Color online)  The carrier spectral density $A^-(k_F,\omega)$ (solid green) for the lower Dirac cone at momentum $k=k_F$ as a function of normalized energy normalized by the chemical potential, $\omega/\mu_0$. The solid black curve gives the interband contribution to the graphene conductivity and its first derivative is given by the dashed blue curve.}
\label{fig10}
\end{figure}

So far we have discuss the intraband conductivity of graphene which is the part most closely related to the conductivity in ordinary one band metals. It is this piece which is most naturally analyzed in terms of an optical self energy and which has been found most useful to obtain information on the spectrum of excitation involved in the inelastic scattering of the charge carriers. But the conductivity in graphene also has a second component which can be nicely separated from its Drude component by increasing the chemical potential $\mu_0$ (charging effects). The interband part in the free band case starts at photon energy $\Omega=2\mu_0$ where it rises sharply to its universal background value $\sigma_0=\pi e^2/(2h)$. When interactions are included the chemical potential shifts to higher energies for the electron-electron case and to lower energies for the electron-phonon and the onset broadens. Here we are interested in the EEI. In this case the rising edge is rather broad as shown by the solid black curve of Fig. \ref{fig10} labeled as interband conductivity and we note modulating structures around $\omega/\mu_0 \sim$ 0.9 and 1.3 coming from the electron-electron correlations. These can be brought out more prominently by taking a first derivative of the interband conductivity which displays two peaks; a narrow, low energy peak and a broad, high energy peak.  These two peaks correlate perfectly with peaks at negative energy in the electron spectral density  for the lower Dirac cone at $k=k_F$, $A^-(k_F,-\omega)$. Transitions from the lower to upper cone at $k=k_F$ appear to dominate the onset of the interband transitions, but the electron-electron interactions clearly provide a substantial broadening of this absorption edge.

\section{Conclusions}

We have calculated the effect of electron-electron (EEI) and electron-phonon (EPI) interactions on the intraband optical effective mass renormalization $\lambda^{op}(\omega)$ and associated optical scattering rate $1/\tau^{op}(\omega)$ of graphene. These quantities are not independent but rather are related by a Kramers-Kronig transformation. They are encoded with information on the effective spectrum of excitations which scatter the massless Dirac fermions responsible for charge transport. For the EEI the quasiparticle self energy is computed in linear approximation in the bare potential reduced by the substrate dielectric function (average over top and bottom dielectrics), which is then screened dynamically in random phase approximation. For the EPI the corresponding self energy follows directly from a knowledge of the electron-phonon spectral density $\alpha^2F(\Omega)$ known in graphene from density functional theory. Once the electronic self energy is known the optical conductivity and optical self energy follows from a Kubo formula. For the simplest case of coupling to a single Einstein boson ($\Omega_E$) mode it is found that the optical mass renormalization exhibits a characteristic easily recognize peak at $\omega=\Omega_E$. This simple observation serves as a qualitative guide for the determination of the distribution of phonons or bosons that may be involved in more complex cases. Examination of our numerical results for $\lambda^{op}(\omega)$ vs $\omega$ when only EEI are considered leads directly to the conclusion that the effective boson spectrum involved in the electron renormalization consists of a relatively uniform distribution extending over an energy range of the order of the chemical potential $\mu$ and scaling linearly with $\mu$. Changing the value of $\mu$ by application of a gate voltage in a field effect device or by some other means such as seeding donor or acceptor atoms on the surface of a graphene sheet, will alter linearly the range of the spectrum while at the same time leaving the value of the spectral $\lambda \equiv 2\int^{\infty}_{0} [\alpha^2F(\omega)/\omega] d\omega$ unchanged.

If in addition to EEI we include also an EPI piece characterized by an Einstein peak for largest contrast, we recover the original EEI background with superimposed, Einstein peak at $\omega=\Omega_E$. While the two distributions cannot be displayed separately, they can still be individually recognized because of their distinct distribution in energy. While we have used for best effect an Einstein peak, a distributed phonon spectrum would still be expected to have sharp peaks and stand out from the relatively unstructured electron-electron background. In addition the shape of the EEI part can be manipulated in a specified way through charging while one would expect the EPI part to be much less affected.

The interband optical transition provide unstructured background to the optical response of graphene distinct and additional to its intraband Drude like response. This new contribution starts at photon energy $\omega=2\mu$ in the bare band case. Correlations smear this rising edge and we find that EEI introduce modulating structures to this edge which are found to correlate remarkably closely with equivalent structures seen in the charge carrier spectral density $A^-(k,-\omega)$ for $k=k_F$ in the lower occupied valence band in the case considered. Another important conclusion of our analysis is that the EEI structures identified in the intraband optical self energy is closely tied to the equivalent structure in the quasiparticle self energy for $k$ equal or close to $k_F$ only. The plasmaronic structure most characteristic of the spectral function $A(k,\omega)$ for $k$ around zero (i.e. around the Dirac point) will display no features in the intraband optical response.

\begin{acknowledgments}

J.H. acknowledges financial support from the National Research Foundation of Korea (NRFK Grant No. 20100008552). J.C. was supported from the Natural Sciences and Engineering Research Council of Canada  and the Canadian Institute for Advanced Research. We thank Prof. E. Schachinger for providing a maximum entropy routine and for his assistance in modifying it to deal with the cases considered here.

\end{acknowledgments}

\bibliographystyle{apsrev4-1}
\bibliography{bib}

\end{document}